\documentclass[12pt]{article}
\usepackage{amsfonts,epsfig,rotating}
\usepackage{cite}

\newcommand{\beq}{\begin{equation}}
\newcommand{\eeq}{\end{equation}  }

\newcommand{\la}{\langle}
\newcommand{\ra}{\rangle}

\newcommand{\bec}{\begin{center}}
\newcommand{\eec}{\end{center}}

\def\nc{n_{\mathrm{c}}}
\def\nt{n_{\mathrm{t}}}

\def\nf{n_{\mathrm{F}}}
\def\nb{n_{\mathrm{B}}}
\def\sf{\sigma_{\mathrm{F}}}

\def\R{\rho}

\def\Nc{\tilde{n}_{\mathrm{c}}}
\def\Nt{\tilde{n}_{\mathrm{t}}}

\def\E{\mbox{e}^+\mbox{e}^-}
\def\pp{\mbox{p}\mbox{p}}
\def\ppb{\mbox{p}\overline{\mbox{p}}}
\def\p+p{\pi^{\pm}\mbox{p}}
\def\K+p{\mbox{K}^{+}\mbox{p}} 
\def\m+p{\mu^{+}\mbox{p}}
\def\np{\nu\mbox{p}}
\def\bnp{\overline{\nu}\mbox{p}}
\def\ep{\mbox{e}^+\mbox{p}}

\begin{document}
\clearpage
\pagestyle{empty}
\setcounter{footnote}{0}\setcounter{page}{0}%
\thispagestyle{empty}\pagestyle{plain}\pagenumbering{arabic}%

\hfill ANL-HEP-PR-98-46 

\hfill July, 1998

\vspace{1.0cm}

\begin{center}

{\Large\bf Long-Range Correlations in Deep-Inelastic \\
Scattering} 

\vspace{2.0cm}

{\large Chekanov S V
\footnote[1]{On leave from
Institute of Physics,  AS of Belarus,
Skaryna av.70, Minsk 220072, Belarus.}
}
 
\vspace{1.0cm}
 
Argonne National Laboratory,
9700 S.Cass Avenue, \\
Argonne, IL 60439
USA

\bigskip

Email address: chekanov@mail.desy.de

\vspace{1.0cm}


\vspace{1.0cm}
   
\begin{abstract}
Multiplicity correlations between the current and target
regions of  the Breit frame in  deep-inelastic scattering processes 
are studied.  It is shown that the correlations are sensitive  
to the first-order  perturbative QCD effects and can
be used to extract the behavior of the boson-gluon fusion rates
as a function of the Bjorken variable.
The behavior  of the correlations is derived analytically 
and analyzed  using a Monte Carlo simulation. 
\end{abstract}

\end{center}
\newpage
\setcounter{page}{1}
\section{Introduction}

Long-range correlations in rapidity  have been 
studied for many years in $\E$, $\m+p$, $\np$, $\bnp$, 
$\pp$, $\ppb$, $\p+p$,
$\K+p$, and $\ep$  collisions in forward-backward
event hemispheres. Results for $\E$, $\m+p$ processes   
indicated  that such correlations are small at 
energies studied in \cite{ee,mp}.
At LEP1 energies, DELPHI, OPAL and ALEPH collaborations
observed  positive long-range correlations, mainly  
due to heavy quark pair production \cite{del,al,op}.   
For $\np$ and $\bnp$ processes, the correlations are 
rather small and negative \cite{np}. 
For $\pp$, $\ppb$ \cite{pp} and  $\p+p$,
$\K+p$ \cite{Kp} collisions, 
the correlations are positive and
increase  with $\sqrt{s}$. Recently, it was  shown  
that the long-range correlations defined in the 
$\gamma^*\mathbb{P}$ center-of-mass system of 
diffractive $\ep$ collisions are positive \cite{ep}.
  
In this paper we discuss the measurements 
of the long-range correlations in neutral current 
deep-inelastic scattering (DIS) $\ep$ collisions using 
the Breit frame \cite{Br}.  We estimate analytically 
the correlations from the
first-order QCD   (Sect.~\ref{sec:qcd}) and compare them 
with a Monte Carlo simulation  (Sect.~\ref{sec:mc}).
 
\section{Definitions}

The event kinematics  of the deep-inelastic processes is 
determined by the 
4-momentum transfer $Q^2=-q^2$, and the Bjorken
scaling variable $x=Q^2/(2 P\, q)$, where $P$ is the 
4-momentum of the proton. 
To study the correlations,  we  use the Breit
frame. In the quark-parton
model (QPM), the  Breit frame  provides the maximum separation between
the radiation from the outgoing 
struck quark and the proton
remnant. In this frame the incident quark carries momentum
$Q/2$ in the positive $z$-direction and the outgoing
struck quark carries $Q/2$ in the negative $z$-direction .
The phase space of the event can be divided into two
regions. All particles with negative $p_z$ components     
of momenta form the current region,
which is analogous to a single hemisphere of $\E$ collisions.
In the QPM, all these particles are produced 
due to the hadronization  of
the struck quark. Particles with positive $p_z$  are
assigned to the target region, which is associated with the 
proton remnant  (see Fig.~1). 


Long-range correlations deal with
the problem of a possible interdependence of different, 
well-separated phase-space
regions of multiparticle production. For the Breit frame, it is
natural to ask  whether the current and target regions are
independent of each other.  
Assuming that a  
color flow between the struck quark and the 
proton remnant at the fragmentation
stage cannot produce strong correlations, 
one may think  that the radiation of the struck quark should be  
independent of the target region.
Below we shall  show that, generally,  
such an expectation is not correct when
QCD corrections to the QPM are considered.
 
To study a possible correlation between the  current and target region,
we  use the  covariance: 
\beq
\R\equiv \mathrm{cov}(\nc , \nt)= \la \nc\, \nt \ra - 
\la \nc \ra \la \nt \ra
\label{1}
\eeq
where $\nc$ ($\nt$) is the number of particles
in the current (target) region, $\la\ldots\ra$ is average over all
events. 

\medskip 
There are a few well established statistical properties of the $\R$:

\begin{enumerate}
\item
Fully independent particle production in the current and
target regions implies $\R=0$.

\item
Since $\R$ represents a degree of linear stochastic dependence
between $\nc$ and $\nt$,  more complex forms of interdependence 
are not described by this variable. 
Therefore, the fact that $\R=0$ is still not
evidence for absence of correlations. 

\item
Positive correlation leads to  $\R>0$,
negative one produces  $\R<0$. 
\end{enumerate}

The forward-backward correlations
have been studied earlier in terms of the parameter $b$  which is
given by the slope in the linear relation $\la \nf\ra =a + b \nb$,
where $\nf$ ($\nb$) is the number of particles in the forward (backward)
hemisphere. The parameter $b$ is directly related to the
covariance $\R$ and can  be defined \cite{eks} using the standard
deviation $\sf$ of the multiplicity distribution in the forward hemisphere,  
\beq
b=\sf^{-2} \R .
\label{ooo}
\eeq

For the present study we shall  use $\R$ rather than $b$. 
While the statistical content of these  quantities is same, 
the  covariance  is easier to  obtain both 
analytically and experimentally.
   
Having established all these notations and definitions, 
below we shall analytically estimate  the covariance $\R$ using first-order
QCD effects.
 

\section{QCD expectation}
\label{sec:qcd}

Going from the quark-parton model to the first-order QCD description 
means to consider more complex processes. In Born approximation, 
Boson-Gluon Fusion (BGF) and QCD-Compton scattering (QCDC) can significantly 
contribute to the overall topology of the events (see Fig.~2). 
Both processes lead to (2+1) partons, where ``1'' denotes 
the proton remnant. While in the QPM  the quark has
$p_z=Q/2$ before and $p_z=-Q/2$ after the collision, 
the first-order QCD processes  make  the collision
in the Breit frame no longer collinear, although 
the target and current regions are still well defined
operationally.   

Following \cite{je}, different event configurations of
(2+1) jets in the Breit frame are shown in Fig.~3.
The configurations are defined according to 
the longitudinal momenta of jets induced by the first-order
QCD processes. For  the  first configuration
both jets from  the first-order QCD processes are moving
to the current region.
For the next two  topologies (see b) and c)),
the two  jets   are produced back-to-back with one jet in the current
and one in the target region. (Both jets may have
different longitudinal momenta, but this does not affect the
QCD estimate of correlations to be made below.)
For the  topology d),  the two  jets are produced in the target region.
Note that, for the Breit frame, the latter topology  is not in conflict
with longitudinal momentum conservation  \cite{je}.

Let now  estimate the covariance $\R$  analytically. 
If $h$  particles migrate to the target region, one can rewrite
(\ref{1}) as
\beq
\R  =
\la (\Nc -h)\, (\Nt +h) \ra - 
\la \Nc -h \ra \la \Nt +h \ra .
\label{r2w}
\eeq
Here  $\Nc$ is the number of particles emitted due to
zero  and first-order QCD processes and 
$\Nt$ is the  multiplicity of the proton remnant without
counting the particles  migrating  
from the first-order QCD processes.
From (\ref{r2w}) one obtains
\beq
\R = \la \Nc\, h \ra - 
\la \Nc \ra  \la h  \ra  -  \la h^2  \ra + \la h  \ra^2 . 
\label{r3w}
\eeq
In this expression, the contributions from the remnant
multiplicity $\Nt$ cancel 
since we consider the case when $\Nt$ is  independent
of  $\Nc$ and $h$. This assumption means that the only dominant
effect leading to the correlation between $\nc$ and $\nt$ is
the first-order QCD migration shown in Fig.~3, 
rather than non-perturbative  effects.
The validity of this assumption will be tested  using a Monte Carlo 
simulation in Sect.~\ref{sec:mc}.

We  do  not  consider   the configuration a) anymore 
since it cannot contribute to the correlation ($h=0$).
This configuration is rather similar to the QPM events
but only with large transverse momenta of both jets. Because of this, it
is possible that some particles
migrate to the target  region  due to the parton showering
or resonance decays.  However, these effects are 
expected to be small.

Now   let us define 
the production rate $R_{b}$ for the back-to-back jets  
(see b) and c) in Fig.~3)  
and the production rate $R_{f}$ for the  events with both
jets moving to the target region
(see d))  as
\beq
R_{b}=\frac{N_{b}}{N_{\mathrm{ev}}}  \qquad
R_{f}=\frac{N_{f}}{N_{\mathrm{ev}}}
\label{r1w}
\eeq
where $N_{b}$ is the number of back-to-back jet events, 
$N_{f}$ is the number of events without  activity in the current region and
$N_{\mathrm{ev}}$ is the total number of events.
We consider  these definitions in the 
limit $N_{\mathrm{ev}}\to\infty$, so that 
$R_{b}$ and $R_{f}$ are the probabilities of having
each configuration.


For every first-order QCD event with a total  multiplicity
$w$, the number $h$ of particles moving to 
the target region is about $w/2$ for the back-to-back jets,
and $w$ for events without  particles in the current region.
Using this estimate, one can write the following set of relations:
\beq
\la \Nc\, h \ra \simeq \la w^2 \ra   \left[0.5 R_{b} + 
R_f \right]
\label{w1}
\eeq

\beq
\la \Nc \ra  \la h  \ra \simeq  \la w \ra \la \Nc \ra \, 
\left[0.5 R_{b} + 
R_f \right]
\label{w2}
\eeq

\beq
\la h^2  \ra \simeq \la w^2  \ra  \left[0.25 R_{b} + 
R_f \right]   
\label{w3}
\eeq

\beq
\la h  \ra^2 \simeq \la w \ra^2  \left[0.5 R_{b}  + 
R_f  \right]^2 .  
\label{w4}
\eeq
Note that  the averaging  for $w$ is performed only  
over the relevant first-order QCD  events.
The term (\ref{w4}) can safely be ignored since $R_{b}, R_f << 1$, 
$\la w \ra^2 < \la w^2 \ra$. 
Combining (\ref{w1})-(\ref{w3}) 
together, one  obtains  from (\ref{r3w})
\beq
\R \simeq  - 
\la w\ra  \la \Nc \ra  \left[0.5 R_{b} +
R_f \right] + 0.25 \, R_{b} \la w^2 \ra . 
\label{w6}
\eeq 

The similarity with $\E$  annihilation makes it possible to estimate
$\la w^2 \ra$.
The distribution of the multiplicity $w$  in the two-jet
events of the $\E$ collisions 
is usually well fitted by  a negative-binomial
distribution  for which the following relation holds
\beq
\la w^2 \ra =\la w \ra^2 (1+k^{-1}) + \la w  \ra 
\label{w7}
\eeq 
with a free parameter $k$ which is much larger than one 
($k=\infty$ for a Poisson distribution). Using this relation,
one obtains 
\beq
\R \simeq - A_b R_b - A_f R_f
\label{w8}
\eeq 

\beq
A_b = \la w \ra  \la \Nc \ra 
\left[ \frac{1}{2} -  
\frac{\la w \ra }{4\la \Nc \ra}(1 + k^{-1}) -
\frac{1}{4 \la \Nc \ra} \right]
\label{w8a}
\eeq 
\beq
A_f =\la w \ra  \la \Nc \ra .
\label{w8b}
\eeq
Below we shall neglect the last term in expression (\ref{w8a}) which is
important only for events at small 
$Q^2$ when $\la \Nc \ra <1$. 

\medskip

The following qualitative predictions can be obtained from 
(\ref{w8}):

\medskip

1) For small $Q^2$, the multiplicity distribution of the QCD induced events
has large $k$, so that it is close to a Poisson distribution. 
Therefore, for similar values of 
$\la \Nc \ra$ and $\la w \ra$, 
the value of $\R$ is negative (anticorrelations). 
Indeed the correlations are  negative if
\beq
\frac{\la w \ra}{\la \Nc \ra} < 2 + 4\, \frac{R_f}{R_b} .
\eeq
This relation  is expected to be a rather good estimate; From $\E$ results
one expects that the  QCD induced events well increase the particle 
multiplicity, however,  the average first-order QCD multiplicity 
$\la w \ra$ cannot be much larger than the 
overall  average multiplicity $\la \Nc \ra$ 
of all events (including the first-order QCD).

\medskip
 
2) The values of $k$ decreases with increasing $Q^2$. This means that
$\R$ increases with energy and even  can change sign. 
The increase with $Q^2$ is mainly determined
by the evolution of the average multiplicities  
$\la w \ra$ and $\la \Nc \ra$ as a function of $Q^2$.

\medskip

3) Let us consider the most interesting case when $Q^2$ is  small and fixed.
Since, for the current region of the 
Breit frame, the evolution of multiplicity distribution 
is determined only by $Q^2$ \cite{lip},  $A_b$ and $A_f$ in (\ref{w8}) 
do not depend on $x$. 
Therefore, (\ref{w8})  has  an  $x$-dependence
determined by  the production  rates   $R_{b}$ and $R_f$. 

Since the  only dominant process at small $Q^2$ is the BGF,
$\R$ in (\ref{w8}) is mainly determined   by the behavior of the
BGF production rate which 
increases with decreasing  $x$ due to increase of the
gluon density inside the proton. Therefore, one can expect
the  magnitude of $\R$ increases with decreasing $x$.

As we see, one of the most striking  features  of 
the Breit frame is a negative value of the long-range correlations.
Such a prediction is rather unusual for  the forward-backward correlations
studied so far. For the DIS  processes in the Breit frame, this
property is quite clear intuitively:  
if one or two jets move  to the target region, then
the fewer particles are observed in the
current region, the more particles can be found in the target region and
vice versa. 

Below we shall  see that 
the analytical observations discussed 
above are in good  agreement with a Monte Carlo simulation.

\section{Monte Carlo study}
\label{sec:mc}

We now  illustrate the  points discussed above using
the LEPTO 6.5 Monte Carlo model \cite{l65}. The model has been tuned as
described in \cite{l65t}. 
The hard process in LEPTO is described by a leading
order matrix element. Below the matrix-element cut-off, parton emission
is based on  the parton shower described by the  
Dokshitzer-Gribov-Lipatov-Altarelli-Parisi
evolution equation.
JETSET Monte Carlo \cite{jetset} based on the
LUND String Fragmentation Model  is used to describe hadronization.      
Although  the production rate of (2+1) jets seems to be underestimated
in LEPTO \cite{wo}, this Monte Carlo model should be 
more adequate for illustrating  
the points discussed above since it is based on an exact first-order 
QCD matrix element calculation.  

To generate DIS  events,
the energy of the positron and  that
of the proton  is  chosen to be
27.5 GeV and  820 GeV, respectively.
We use the following cuts 
$$
Q^2 > 10\> \mbox{GeV} \qquad y \le 0.95 \qquad E\ge 10\> \mbox{GeV} 
$$
where $y$ is the relative energy transfered from the electron to
the proton in the proton rest frame, $E$ is the energy of
scattered electron. In total,  200k events are  generated for
each  of the measurements to be discussed below. 


\subsection{First-Order QCD Rates}

Let us  first consider  the first-order QCD rate for the BGF and QCDC.
We define the production rates $R_{\mathrm{BGF}}$ and 
$R_{\mathrm{QCDC}}$ for the BGF  and QCDC, respectively, as
\beq
R_{\mathrm{BGF}}=\frac{N_{\mathrm{BGF}}}{N_{\mathrm{ev}}}
\qquad R_{\mathrm{QCDC}}=\frac{N_{\mathrm{QCDC}}}{N_{\mathrm{ev}}}
\label{r1}
\eeq
where $N_{\mathrm{BGF}}$ and $N_{\mathrm{QCDC}}$ are the numbers of the
BGF and QCDC events.  Experimentally,
the information about sum of these rates  can be obtained from the study
of  (2+1) jet rate $R_{2+1}$ \cite{h1,zeus}.  Note that 
$R_{\mathrm{BGF}}+R_{\mathrm{QCDC}}$ is not
equivalent to  $R_{2+1}$.  The latter depends on the 
jet algorithm and a resolution scale
$y_{\mathrm{cut}}$  to define jets. Note that the normalizations in
$R_{2+1}$ and  $(\ref{r1})$ can  also be different.

For the Monte Carlo study, (\ref{r1}) can directly be obtained
by counting the first-order QCD events\footnote{
The parameter LST(24) specifies the type of the first-order QCD event
in LEPTO model.}.
The production rates in LEPTO are  derived
numerically as the integral of the relevant first-order matrix elements.
They depend on cut-offs  on the matrix elements and involve the GRV94 
parton density parameterization \cite{pdf}. 

Fig.~4 shows the production rates  of 
BGF and  QCDC as a function of $Q^2$ and 
$x$. For the latter figure,
the cut $Q^2 \le 50 \> \mbox{GeV}^2$ is used to constrain the effect
of increase of  $Q^2$ with increasing  $x$.

For $Q^2$ variable, the BGF rate rises 
with increasing $Q^2$ for $Q^2 < 100$ $\mathrm{GeV}^2$,
and then it falls. 
For $x$ variable, the BGF rate increases  with decreasing $x$.
Since for all range of $x$  studied 
the average value of $Q^2$ only increases from $\sim 19$ $\mathrm{GeV}^2$ 
(for  $\la x\ra \sim 0.0007$) to $21$ $\mathrm{GeV}^2$ 
(for $\la x\ra \sim 0.02$),
such a behavior  is mainly because of the variations of $x$ 
due to an increase of the gluon density inside proton.

\subsection{Current-Target Multiplicity Correlations}

Fig.~5 shows the behavior of covariance  $\R$
as a function of $Q^2$ and $x$. 
The values of $\R$ for the QPM  are  near  zero, 
i.e. there is no strong linear interdependence between
the current and target regions\footnote {Note again that, generally, 
a non-linear interdependence may exist.}.  
This illustrates the fact that the LUND String Fragmentation Model 
used by LEPTO does not produce large  
correlations during the formation and an independent breaking of strings 
stretched between the  current-region showering partons and the remnants.
This is further illustrated in \cite{ccc}.  

Adding BGF and QCDC events
leads to the negative correlations  expected. 
The magnitude  of $\R$  rises  with increasing $Q^2$ 
(for $Q^2 < 100 \> \mathrm{GeV}^2$) and then falls.
Such a behavior is due to the similar trend of the BGF rate shown
in Fig.~4.
Note that a decrease  of $k$  
in (\ref{w8}) can also contribute to the decrease of the correlation. 
For $Q^2 < 100 \> \mathrm{GeV}^2$, 
symbols for QPM+BGF and the LEPTO default (QCD)
are very close to each other.

The  dependence of $\R$ as a function of $x$ is mainly 
due to the variation of $x$, since there are  no large variations in
$Q^2$ for the cuts applied.  The magnitude  of $\R$  increases 
with decreasing  $x$. 
Symbols for the QPM with  all first-order QCD effects 
(QPM+BGF+QCDC) and for QPM+BGF 
events are very close, i.e. the boson-gluon fusion is
the main source of the correlation. 
For larger $x$,  symbols become
distinguishable  since the contribution of the BGF  becomes smaller  at
large $x$. 

According to analytical expression (\ref{w8}),
there exists  a linear relationship between the $\R$ 
and probabilities $R_b$ and $R_f$ determining the
BGF production rate. 
The solid line shown in Fig.~5  illustrates  the BGF rate 
(see  Fig.~4) multiplied by  the scale factor $-6.5$.
The shaded band shows the statistical errors for BGF rate.
As seen, the behavior of $\R$ follows  that of  
BGF rate rather well.

Note that for very small $x$, one has to expect a deviation from 
the linear relationship between $\R$ and BGF rate since 
the non-linear term omitted in (\ref{w6}) cannot be longer neglected.


\section{Conclusion}

Long-range correlations in DIS were  investigated
using the Breit frame.
It was shown  that the correlations  between the
current and target regions of  the Breit frame 
are   sensitive to the first-order QCD processes leading to
the (2+1) jets.
In particular, at small $x$ and a restricted interval of $Q^2$,
the strength of the long-range correlations  is mainly determined 
by the boson-gluon fusion process. One of the distinguishing
properties of the correlations  is their negative value and rise in  the
magnitude with decreasing $x$.

This method can be used to study the production rates
of the boson-gluon fusion process,  
to understand better the gluon density inside proton
and to discriminate between different Monte Carlo models used to simulate the
deep-inelastic processes. 
The approach has the advantage that it is simple and does not involve any 
jet algorithm and  a resolution 
scale necessary to  determine jet, since it is based  on purely 
topological properties of (2+1) jets  in the Breit frame. 
Therefore, the advantage of this method is absence of the
systematical uncertainties  connected
with the  ambiguity in  determining  jets. 

Note also that the  jet definitions are not able to distinguish between the 
proton remnant and QCD induced jets if they are close to each other.
This problem is avoided in the proposed jet-rate measurement:
The strength of the correlations is determined on the basis of an  
enhancement of the  multiplicity in the target region for
the first-order QCD events, rather than resolving well-separated 
jets.

The current-target correlations can be used to study a difference
in particle spectra  
between the current region of DIS  and 
a single hemisphere of $\E$ interactions. 
In contrast to DIS processes exhibiting
anticorrelations in the Breit frame, 
the  forward-backward correlations between two opposite rapidity 
hemispheres of $\E$  annihilation
are positive \cite{del,al,op}. This 
may  lead  to discrepancies in attempts to compare the current
region of DIS  with a 
single  hemisphere of $\E$ interactions.

Estimating the current-target multiplicity correlations
analytically, we did not take into account  possible color exchange
effects between the fragmentation of the 
outgoing partons and that of the remnant.
Also we did not discuss  
high-order QCD corrections.
These topics  are  rather important for a quantitative confrontation of the
correlations with the data and have to be studied in the future.

\bigskip
\section{Acknowledgments} 
\medskip 

I thank Derrick~M, De~Wolf~E,  Doyle~T,   L\"{o}nnblad~L,  Magill~S,  
Repond~J  for valuable  discussions. 

{}

\newpage

{\Large\bf Figure captions }

\vspace{1.0cm}

\begin{itemize}

\item
{\bf Figure 1}: 
a) Diagram for the neutral current deep-inelastic
scattering in the QPM;
b) A schematic  representation of the Breit frame.
Particles with $p_z<0$ belong to the current region.
Particles with $p_z>0$ form the target region.

\item
{\bf  Figure 2}: 
Diagrams in the first-order perturbative QCD:
a) Boson-Gluon Fusion (BGF);  b) QCD Compton scattering (QCDC).
For the latter diagram
the gluon can also be radiated before the interaction with
$\gamma^*$.

\item
{\bf Figure 3}: 
Typical configurations of (2+1) jets in the Breit frame \cite{je}.
''1'' denotes the proton remnant, ''2'' and ''3'' are
jets due to the first-order QCD corrections to the parton model.
Jets with $p_z<0$ form the current region. For $p_z>0$,
they  belong to the target region.

\item
{\bf Figure 4}:
Production rates for the BGF and QCDC as a function
of $Q^2$ and $x$.

\item
{\bf Figure 5}:
The values of the covariance $\R$  for different bins
in $\la Q^2 \ra$ and $\la x \ra$ obtained from  the LEPTO 6.5
Monte Carlo model.
We show four event topologies:
1) quark-parton model (QPM); 2)  QPM with the
first-order QCD events (QPM+BGF+QCDC)
3) QPM with BGF events; 4) QPM with QCDC events.
For the latter figure, symbols
for QPM+BGF+QCDC and QPM+BGF are on top of each other.
\end{itemize}

\newpage
\begin{figure}
 
\vspace{2.0cm}
\begin{center}
\begin{sideways}
\begin{sideways}
\begin{sideways}
\mbox{\epsfig{file=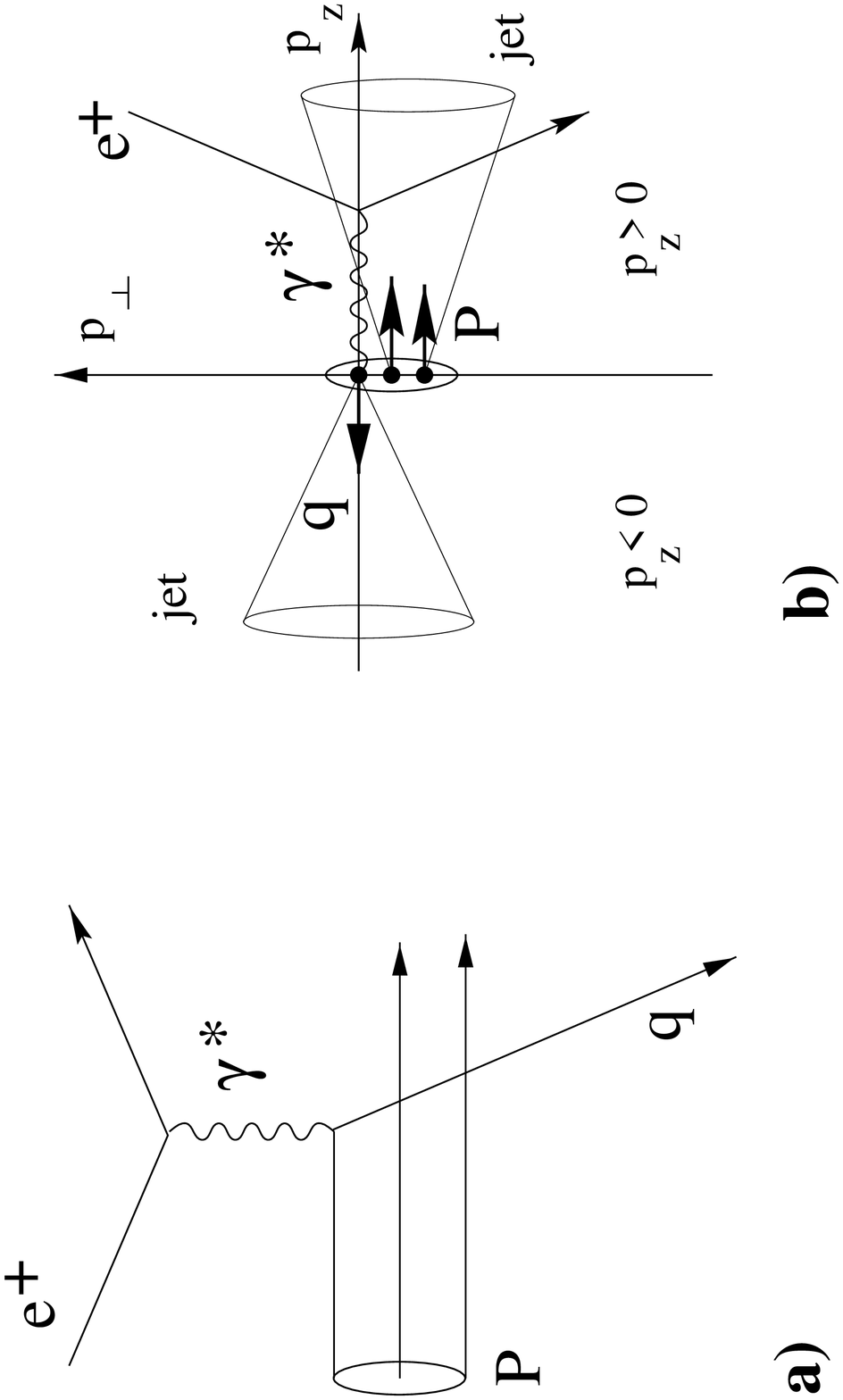,height=15.0cm}}
\end{sideways}
\end{sideways}
\end{sideways}
\end{center}

\vspace {2.0cm}
\centerline{\large\bf Figure 1}
\end{figure}

\newpage 
\begin{figure}
\begin{center}
\vspace{1.5cm}
\begin{sideways}
\begin{sideways}
\begin{sideways}
\mbox{\epsfig{file=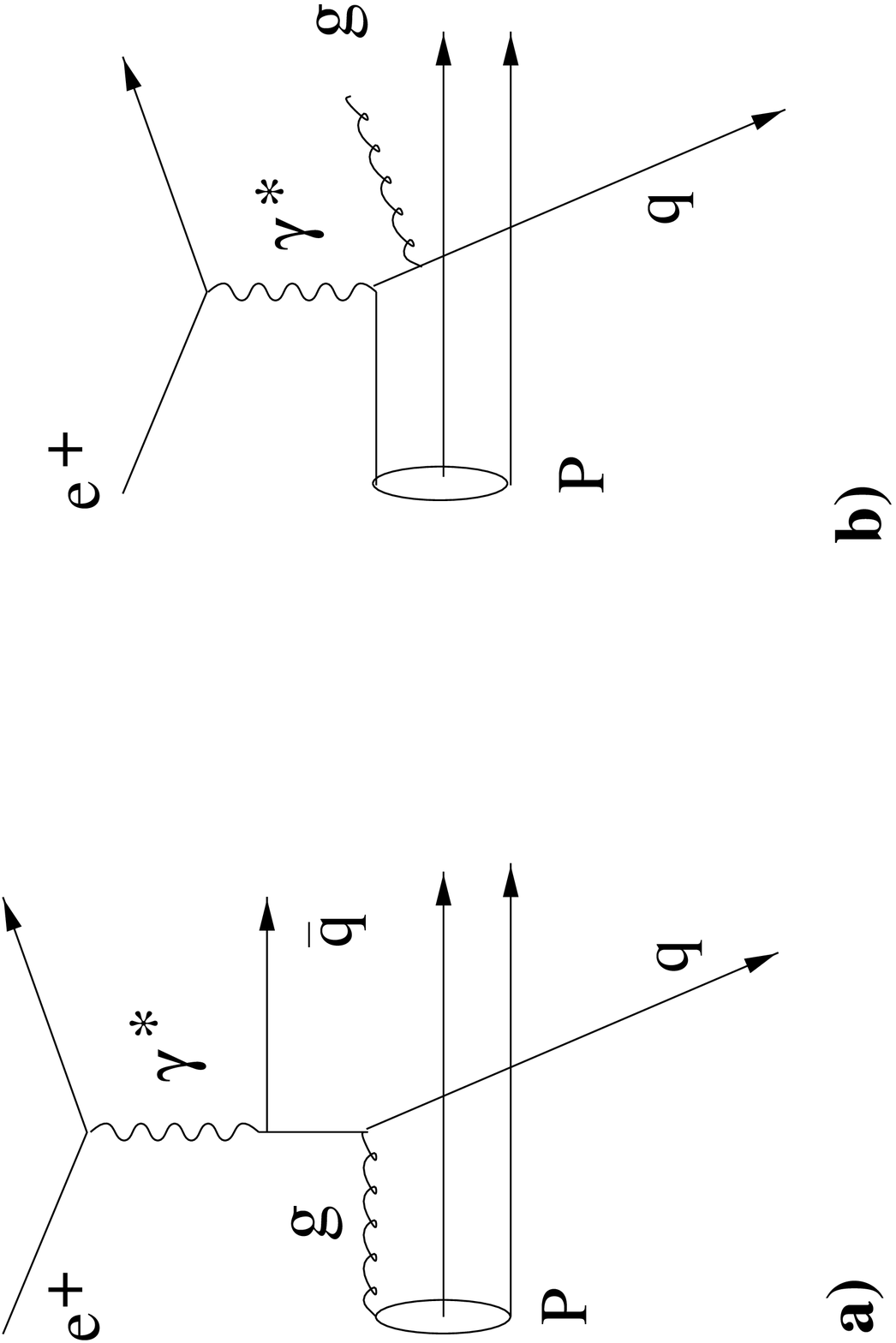,height=9.0cm}}
\end{sideways}
\end{sideways}
\end{sideways}
\end{center}

\vspace {1.5cm}
\centerline{\large\bf Figure 2}
\end{figure}

\newpage 
\begin{figure}
\begin{center}
\vspace{1.0cm}
\begin{sideways}
\begin{sideways}
\begin{sideways}
\mbox{\epsfig{file=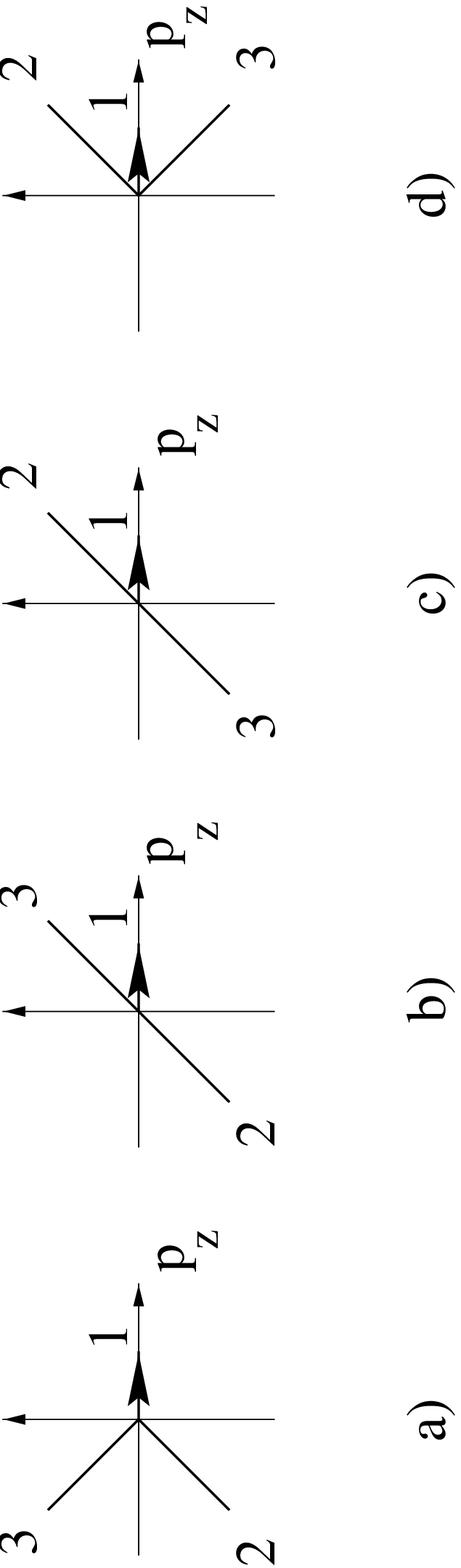,height=10.0cm}}
\end{sideways}
\end{sideways}
\end{sideways}
\end{center}

\vspace {1.5cm}
\centerline{\large\bf Figure 3}
\end{figure}

\newpage
\begin{figure}
\begin{center}
\mbox{\epsfig{file=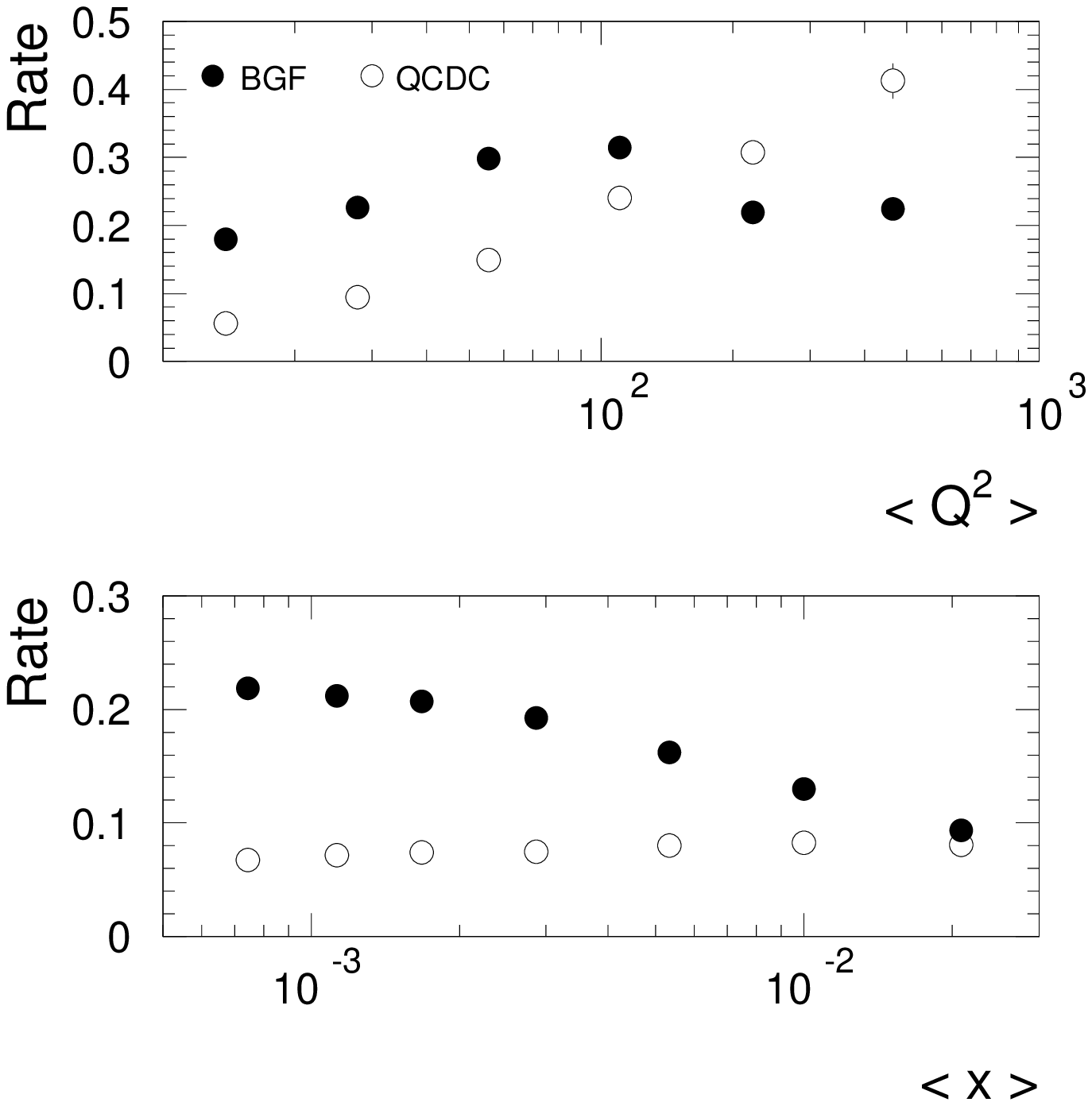,height=18.0cm}}
\end{center}

\vspace {1.5cm}
\centerline{\large\bf Figure 4}
\end{figure}

\newpage 
\begin{figure}
\begin{center}
\mbox{\epsfig{file=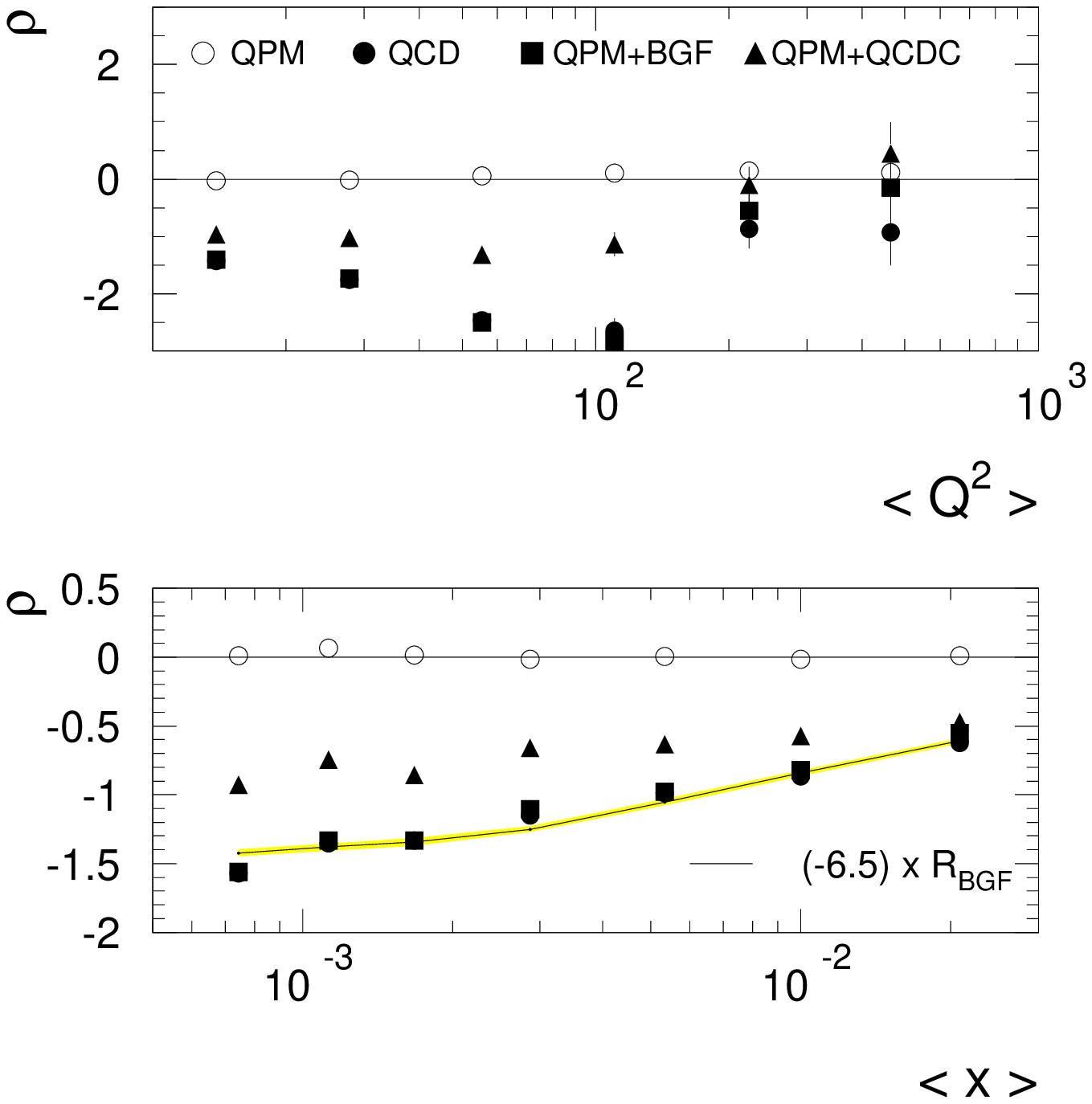,height=18.0cm}}
\end{center}

\vspace {1.5cm}
\centerline{\large\bf Figure 5}
\end{figure}

\end{document}